\documentclass[conference]{IEEEtran}
\IEEEoverridecommandlockouts
\usepackage{comment}
\usepackage{cite}
\usepackage{amsmath,amssymb,amsfonts}
\usepackage{algorithmic}
\usepackage{graphicx}
\usepackage{textcomp}
\usepackage{xcolor}
\usepackage{subcaption}
\usepackage[a4paper, total={184mm,239mm}]{geometry}
\def\BibTeX{{\rm B\kern-.05em{\sc i\kern-.025em b}\kern-.08em
    T\kern-.1667em\lower.7ex\hbox{E}\kern-.125emX}}

\usepackage{listings}

\usepackage{fancyhdr}

\usepackage{pifont}
\newcommand{\cmark}{\ding{51}}%
\newcommand{\xmark}{\ding{55}}

\usepackage{fancyhdr}
\makeatletter
\newcommand*\titleheader[1]{\gdef\@titleheader{#1}}
\AtBeginDocument{%
  \let\st@red@title\@title
  \def\@title{%
    \bgroup\normalfont\large\centering\@titleheader\par\egroup
    \vskip1.5em\st@red@title}
}
\makeatother

\title{A Parallel SystemC Virtual Platform for Neuromorphic Architectures\vspace{-5mm}}
\titleheader{\vspace{-10mm}\textcolor{red}{Preprint - Accepted at 23rd International Symposium on Quality Electronic Design (ISQED'22)}\vspace{-10mm}}

\author{\IEEEauthorblockN{Melvin Galicia, Farhad Merchant  and Rainer Leupers}
\IEEEauthorblockA{\textit{Institute for Communication Technologies and Embedded Systems} \\
\textit{RWTH Aachen University}\\
Aachen, Germany \\
\{galicia, farhad.merchant, leupers\}@ice.rwth-aachen.de}
\vspace{-5mm}}

\begin{document}
\bstctlcite{IEEEexample:BSTcontrol}
\maketitle
\begin{abstract}
With the increasing interest in neuromorphic computing, designers of embedded systems face the challenge of efficiently simulating such platforms to enable architecture design exploration early in the development cycle. Executing artificial neural network applications on neuromorphic systems which are being simulated on virtual platforms (VPs) is an extremely demanding computational task. Nevertheless, it is a vital benchmarking task for comparing different possible architectures. Therefore, exploiting the multicore capabilities of the VP's host system is essential to achieve faster simulations. Hence, this paper presents a parallel SystemC based VP for RISC-V multicore platforms integrating multiple computing-in-memory neuromorphic accelerators. In this paper, different VP segmentation architectures are explored for the integration of neuromorphic accelerators and are shown their corresponding speedup simulations compared to conventional sequential SystemC execution. 
\end{abstract}

\section{Introduction}
The limitations of the von Neumann model have forced researchers to look into alternate computer architectures. The von Neumann model fares poorly due to the energy spent in data movements especially for machine learning applications in embedded edge devices, where a huge amount of unstructured data is to be processed~\cite{neurosurvey2}. Emerging non-volatile memory technology such as resistive random access memory (ReRAM) offers a unique advantage in-terms of computing-in-memory features along with storage capabilities~\cite{Lu1}. Such a feature alleviates the need for data movement between a processor and memory subsystem significantly, resulting in energy efficiency and high-performance of a neuromorphic architecture. The design and development of a neuromorphic platform requires careful consideration of several architectural parameters~\cite{reram1}. In addition, the software's performance to be executed needs to be tuned to attain the best possible performance at an early stage in the design cycle. 

A virtual platform (VP) offers the simulation of architectural components to achieve the right balance between time to market and the performance of the target product~\cite{vp2}. Especially for complex platforms, VPs offer a unique advantage that the software development can begin significantly earlier than the hardware prototype's availability~\cite{vp1}.    

The emergence of neuromorphic computing-in-memory (CIM) architectures in the last decade has propelled research to develop various simulation platforms for performance prediction and early-stage software development. While some of these simulators lack the required accuracy, others are extremely slow due to the details involved in the simulation. Additionally, the application benchmarks used in machine learning are extremely demanding computational tasks, which mainly require vector-matrix multiplication (VMM) operations that also have to be executed within the VP. All this creates the need of a VP capable  of effectively using all computing power available to its host system. So, it can speed up the simulations required to perform investigation of neuromorphic architectures. To address these issues related to neuromorphic VPs, this work introduces a SystemC based VP that utilizes, since its conception, parallelization techniques to efficiently exploit all available CPU resources in the host system. The major contributions of this paper are as follows:
\begin{itemize}
\item A complete full system-level neuromorphic VP able to run a heavy load of VMM benchmarks in a  parallel simulation environment.
\item Simulation segmentation strategies to distribute simulation load among multiple cores of the host system in a comprehensive method.
\item Utilization of time-decoupled parallelization techniques to further enhance the VP's simulation speedup.
\end{itemize}

This paper is structured as follows: Section II provides related work and background. Section III introduces the structure of the VP and its characteristics. Section IV summarizes the used time-decoupled and parallel SystemC methodologies. In section V, the significance of time-decoupled parallel SystemC VP is demonstrated by using it to perform two different VP segmentation strategies of multicore and multi-CIM systems. It also presents benchmark results versus sequential simulations for all analyzed cases. Finally, in section VI conclusions are drawn and an outlook of future research is provided.

\section{Related Work}
\begin{figure*}[!t]
\centerline{\includegraphics[scale=0.45]{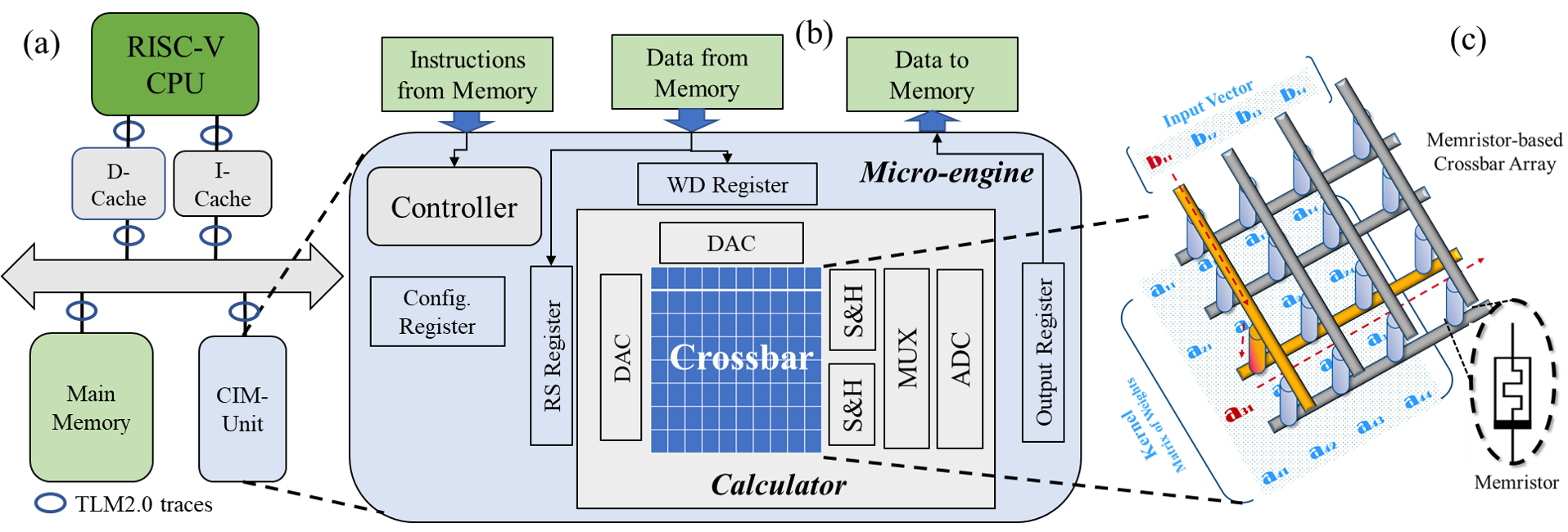}}
\vspace{-0mm}
\caption{Single-core System-level virtual platform. (a) VP high level architecture, (b) Neuromorphic computing-in-memory architecture: CIM-Unit, (c) Memristor-based crossbar array.}
\vspace{-3mm} 
\label{figX}
\end{figure*}

\begin{table*}[!h]
\scriptsize
\centering
\caption{Qualitative comparison of the simulators in the literature}
\vspace{-0mm}
\begin{tabular}{|p{1.8cm}| p{1.2cm} | p{1.2cm}| p{1.2cm}| p{1.1cm}| p{1.7cm}| p{0.8cm}| p{1.7cm}| p{2.0cm}|}
\hline
Simulator & Architecture-level & System-level  & Circuit-level & Exploration possibility  & Parallelization & CIM Support & Accelerator-enabled & Time decoupling-enabled \\
\hline
NVSim~\cite{nvsim}      & \xmark & \xmark & \cmark & \cmark & \xmark & \xmark & \xmark & \xmark  \\ \hline
NVMain~\cite{nvmain1}       & \cmark & \xmark & \xmark & \cmark & \xmark & \xmark & \xmark & \xmark  \\ \hline
MNSim~\cite{mnsim1}     & \cmark & \xmark & \xmark & \xmark & \xmark & \cmark & \xmark & \xmark  \\ \hline
CorssSim~\cite{crosssim}     & \xmark & \xmark & \cmark & \cmark & \xmark & \cmark & \xmark & \xmark  \\ \hline
NeuroSim~\cite{neurosim}  & \xmark & \xmark & \cmark & \cmark & \xmark & \cmark & \xmark & \xmark  \\ \hline
NVM-SPICE~\cite{nvmspice}  & \xmark & \xmark & \cmark & \cmark & \xmark & \xmark & \xmark & \xmark  \\ \hline
AIHWKIT~\cite{aihwkit}  & \xmark & \xmark & \cmark & \cmark & \xmark & \cmark & \xmark & \xmark  \\ \hline
CIMSIM~\cite{CIM-SIM}       & \xmark & \cmark & \cmark & \xmark & \cmark & \cmark & \xmark & \xmark  \\ \hline
Lee et al.~\cite{Lee2019} & \xmark & \cmark & \xmark & \cmark & \xmark & \cmark & \xmark & \xmark  \\ \hline
MultiPULPly~\cite{TechnionSystemL}  & \xmark & \cmark & \xmark & \cmark & \cmark & \cmark & \xmark & \xmark  \\ \hline
Proposed work  & \cmark & \cmark & \cmark & \cmark & \cmark & \cmark & \cmark & \cmark  \\ \hline
\end{tabular}
 \label{tab:neuro_literature}
 \vspace{-5mm}
\end{table*}

With the advancements in emerging non-volatile memory technologies, various research groups have tried to explore the technologies at the system level. On the other hand, for the CMOS-based electronic systems, SystemC-based virtual prototypes are extremely popular for early software development and performance tuning. In this section, we first review various simulators and then briefly discuss the relevant systemC-based virtual prototypes.  

A system-level simulation of a tiled neuromorphic architecture is presented in~\cite{CIM-SIM}. The authors have implemented a micro-instruction set to perform the operations on analog/digital memristive crossbars. Along with the crossbars, each tile consists of analog/digital converters, digital modulators, and sample-hold mechanisms. The micro-instruction has three types of functions: initialize, compute, and read. The system-level simulator presented in~\cite{Lee2019} is built on similar principles. NVMain~\cite{nvmain1} and NVMain 2.0~\cite{nvmain2} are the simulators that can simulate memory and memory interfaces. NVMain focuses on memory-oriented simulations and requires significant engineering efforts in modeling and simulation of CIM. 

The MNSIM~\cite{mnsim1} and MNSIM 2.0~\cite{mnsim2} simulators use behavioral models to estimate the worst case and accuracy. The increased simulation speed comes at the cost of accuracy in MNSIM. Apart from the system level and architecture simulators, there are various simulators capable of performing circuit-level simulations~\cite{nvsim}~\cite{neurosim}. While these simulators can perform low-level simulations and provide better estimates of the energy efficiency of the design, scaling to system-level is not possible due to simulation complexity and run-time.

VPs commonly use SystemC as their underlying simulation engine, which deploys a standard discrete event approach to simulate the concurrency of multiprocessor systems. There are several approaches for parallelization of SystemC \cite{ParalleSysC1,OoOparallelsym,LegaSCI}. A synchronous approach to parallel SystemC simulation is presented in \cite{LegaSCI}. Synchronous techniques, however, are unable to exploit the full parallelism of a simulation, given that they enforce strict time synchronization among all threads. This synchronous limitation can be overcome by applying a time-decoupled technique as proposed in \cite{timedecoup}. These techniques have been applied in \cite{SysCArm} to multicore VPs for specific workloads like booting operating systems and for network intensive communication applications. However for heavy VMM operations needed in machine learning applications, that are executed in a system-level VP with multiple neuromorphic accelerators architecture, there is limited work available. Extensive work on different simulators for neuromorphic architectures with different abstraction levels is summarized in~\cite{neurosurvey2}, where the majority of simulators focus on circuit level. Only a few works like~\cite{ICESOCC,TechnionSystemL} have presented complete integration of neuromorphic accelerators into a full system. However, they have not been explicitly developed to exploit the multicore nature of their host systems. Therefore, this work presents a time-decoupled parallel simulation framework that allows different VP segmentation strategies of neuromorphic architectures. A summary for several neuromorphic simulation frameworks is captured in Table~\ref{tab:neuro_literature} where a qualitative comparison can be observed. The proposed simulation framework has a few features that are desired but absent in simulators found in the literature.

\section{Virtual Platform}	
The proposed system-level VP consists of several SystemC modules: a set of multicore processors, L1 instruction and data caches, communication buses, a main memory module and multiple memristor-based CIM neuromorphic accelerators. Each SystemC module can work as a transaction initiator and a transaction target depending on its current function or just as a forwarder in the case of the bus modules. The modules have their corresponding sockets for TLM2.0 transaction-based communication where a tracing monitoring of each transaction is being performed and stored for post-processing performance analysis of the whole system. Fig.~\ref{figX}a depicts the VP high level architecture and illustrates the tracing approach. By collecting all the transactions, it is possible to have a histogram and the total number of transactions that occurred during the execution of any particular benchmark.
\subsection{RISC-V}	
The VP uses RISC-V \cite{Waterman:EECS-2014-54}, an open-source instruction set architecture, as the main processor of the multicore system. The RISC-V core implemented in \cite{HERDT2020101756} using SystemC and TLM-2.0 is the one integrated into the VP. This implementation offers a 32/64-bit RISC-V core supporting the IMAC instruction set. By adding this SystemC RISC-V 64IMAC, the VP achieves a significantly faster simulation compared to using an RTL implementation. The RISC-V processor is used to execute different software applications that exploit the available neuromorphic accelerator by offloading artificial neural network (ANN) main operations, i.e. VMMs, meanwhile being free to perform other computational tasks when necessary. It is important to point out that the RISC-V core does not make use of any standard extension instructions to offload operations to the neuromorphic accelerator. It is the CIM unit that interprets and handles the executions of the operations using its internal controller and own set of micro-instructions as explained in next section.

\subsection{Memristor-Based CIM-Unit}
The VP integrates multiple CIM-Units of the one proposed in \cite{CIM-SIM}, since it provides built-in interfaces that allow interaction with SystemC based VPs in several abstraction levels. Each CIM-Unit consists of two main parts: A calculator with its digital surrounding and a micro-engine as shown in Fig.~\ref{figX}b. The calculator is a simulator that not only replicates the functional behavior of a memristor crossbar shown in Fig.~\ref{figX}c, but also covers the operation of surrounding mixed-signal circuitries, e.g., analog to digital converter (ADC), digital to analog converter (DAC) and sample and hold (S+H) elements, which are essential for driving the memristor crossbar. The micro-engine comprises the digital components, e.g., controller, registers files, and buffers, that are necessary for operating the memristor crossbar. In addition to these, the CIM-Unit offers a micro-instruction set that allows the unit to communicate with the main processor or other functional units, in a coarse grain reconfigurable architecture. Thus, it can be deployed as an accelerator as it is the case in this work. For the CIM-Unit to operate it needs first be supplied with configuration parameters, e.g., size of the matrix that is to be mapped on the crossbar (a fixed size of 256x256 was configured), I/O resolution as well as others necessary parameters detailed in \cite{CIM-SIM}. The controller stores all these parameters which then are passed to the CIM-Unit through instructions via the configuration register. As soon as all the parameters are available, the controller, which is a state-machine, transitions into the state IN signaling that is ready to accept input data. It stays in IN until all the input data is received. The number of cycles that it takes to be ready to move to the next state depends on the width of the input ports, the resolution of the input data, and the size of the input vector. In the next state, OP, the controller executes the operation that is specified in the configuration register. In the end, the controller moves to the final state, OUT, where it sends out the processed information and goes back to the initial state, IDLE, as soon as all the output is sent out.

\subsection{Memory Modules}
The main memory (DRAM), which works as a shared module for all segments, and instruction and data cache (SRAM) modules, from the work presented in \cite{memCache}, are the ones integrated in the VP. Since in addition to commonly modeling memory in SystemC with a read and write access delay, further delays for page switches and write-to-read switches are introduced.

\section{Parallel SystemC Simulation}
This work implements time-decoupled parallel simulation using techniques investigated in \cite{sysClink}, and also presents different segmentation methodologies. The segmentation process can be used to explore effective exploitation configurations for parallel execution of VMM benchmarks in a multicore system integrating multiple neuromorphic accelerators. Aforementioned implementation and methodologies are briefly summarized as follows:

\begin{figure}[!t]
\centerline{\includegraphics[width=0.35\textwidth]{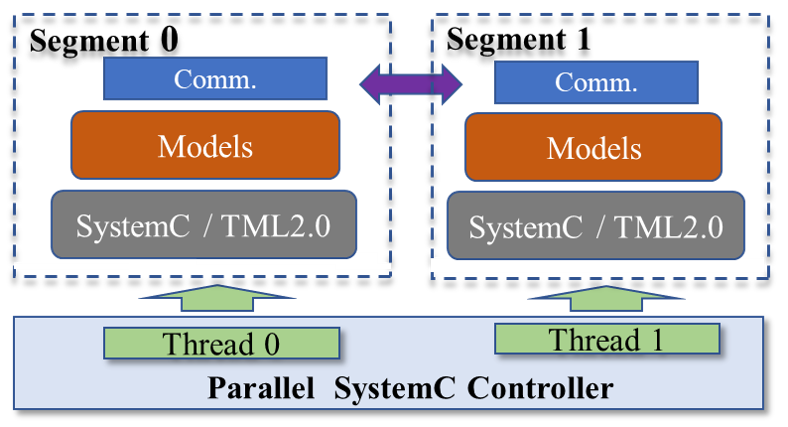}}
\vspace{-2mm}
\caption{SystemC parallel simulation overview.}
\vspace{-1mm}
\label{figSC_P}
\end{figure}

\begin{figure*}[!t]
\centerline{\includegraphics[width=1\textwidth,height=3cm]{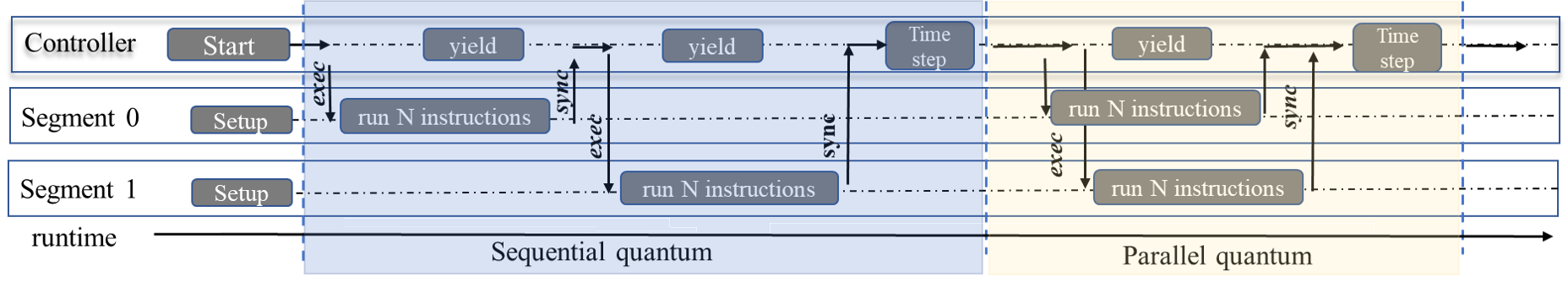}}
\vspace{-3mm}
\caption{SytstemC sequential and time-decoupled parallel simulation quantum.}
\vspace{-5mm}
\label{figquantum}
\end{figure*}

\subsection{Parallel Simulation Architecture}
A complete simulation consists of \textbf{\textit{n}} segments, defined during the segmentation process, that are interconnected via dedicated communications channels. By organizing the simulation as a list of segments each containing its own SystemC kernel, parallel simulation becomes simple, since it is not necessary to modify the internal discrete event simulation of each kernel. As a result, each segment can be stepped independently and is then parallelized by the host system. However, communication, time-decoupling and post-synchronization become key procedures during the whole simulation execution. Therefore, the  simulation controller is used to advance the simulation by stepping each segment and enforcing the sync mechanism. The simulation flow that executes SystemC for the simulation segments in parallel is shown in  Fig.~\ref{figSC_P}.

\subsection{Time-Decoupled Simulation}

Higher simulation speed can be achieved by deploying time-decoupling between the segments \cite{timedecoup}. This is achieved by relaxing the synchronization requirement between segments and allowing the simulation controller to let the simulation run in parallel for a certain time before synchronization is enforced. The controller organises instructions into quanta: a chunk of N instructions that are grouped together and executed, before simulation time is advanced by a corresponding amount. The threads runtime are constructed based on three communication primitives: \textbf{\textit{setup}}: loads the thread and sets quantum length to N,\textbf{\textit{ exec}}: asynchronously executes the next N instructions, \textbf{\textit{sync}}: blocks until the thread has finished its current quantum. This is illustrated in Fig.~\ref{figquantum} for both cases, when parallel execution is performed and when it is done sequentially by default in SystemC.

Since each segment now operates in its own time zone with a local time, inter-segment communication is assigned a latency. Channel latency describes the amount of time allowed to pass after putting a message into a channel before it must be fetched by the receiver. It is used by the simulation controller to determine the amount of time a segment may simulate ahead of time before risking missing a channel message from a peer segment. Hence, it is the task of the simulation controller to make sure that no segment has advanced too far ahead in time. Segments that have not reached their limit time are considered in simulation, while others are considered waiting for their peers to match time.

\subsection{Segmentation}
The segmentation process allows the designer to explore different partition strategies of the VP  architecture for the integration of neuromorphic CIM-Units within a multicore system. There are several parameters that can be considered to build a complete system, e.g. the total number of CPUs, the total  number of CIM-Units and also the division of load execution. In Fig.~\ref{figSg1} the segmentation corresponds to an equal number of CIM-Units accessible to each CPU, i.e. uniform VP segmentation. It is as well possible to segment in a way that one CPU is dedicated to manage all CIM-Units while other CPUs are free to perform other tasks, i.e. load-oriented VP segmentation as shown in Fig.~\ref{figSg2}. Subsequently, the CIM-Units can be allocated in parallel segments to speed up even further the simulation.
Segments form virtual sequential environments which are allowed to share data using communication channels. Segments are implemented as shared object files with unique names. Hence, the system can easily be extended to a \textbf{\textit{n}}-core system by instantiating  more segments and creating new channel connections. Once the simulation starts, the controller selects the first segment from the ready to simulate queue and starts its execution each in its corresponding thread.

\section{Experimental Evaluation}
To demonstrate the efficacy of the proposed approach, it is evaluated in various experiments. The evaluation considers VMM loads from well known ANN layers to assess the performance benefits of the parallel simulation approach. Runtimes for all workloads are analyzed using the conventional sequential (sq) and the parallel simulation approach (pll).

\subsection{Multicore and Multi-CIM Systems}
The designs of the VP with uniform and load-oriented segmentation, shown in Fig.~\ref{figSg1} and  Fig.~\ref{figSg2} respectively, are explored in this work. In the first case only two segments are initialized, each is allocated with one RISC-V processor and with only two CIM-Units. In the second case, four segments are initialized, segment 0 contains a RISC-V processor and the main shared memory. Segment 1 contains only one RISC-V processor, meanwhile segment 2 and 3 each contains only two CIM-Units. The VP components and their parameters are listed in Table ~\ref{tabcomp} as well as for the host machine.

\subsection{Benchmarks}
Convolutional layers from Googlenet \cite{googlenet}, ImageNet \cite{Imagenet}, and MobileNets \cite{mobilenets} ANNs are selected as benchmarks that run on the VP, these are heavy workloads that require acceleration for the vector matrix multiplication operations. Due to the multicore nature of the simulated VP, it is possible to run the convolutional layers using one CPU and main memory as well as running them by offloading the operation to the available CIM-Units.  The layers are listed in Table~\ref{tabbench}, where \textbf{\textit{h}} and \textbf{\textit{w}} are the matrix height and width respectively and \textbf{\textit{p}} is the number of vectors. The matrix and vector sizes varied to assess different scenarios. Nonetheless, all matrices fit into the configured crossbar inside the CIM-Units. A simple nested loop algorithm for a VMM operations was implemented to obtain the \textbf{O} matrix which is the dot product between matrices \textbf{A} and \textbf{B}. The algorithm was used for the case of RISC-V plus the shared main memory and also for the case when the host system offloads the operations to the CIM-Units. However for the latest case, the internal operation handling is left to the micro-engine component of the CIM-Units. The nested loop algorithm is described below:



\begin{figure*}[!t]
     \centering
     \begin{subfigure}[b]{0.45\textwidth}
         \centering
         \includegraphics[width=\textwidth]{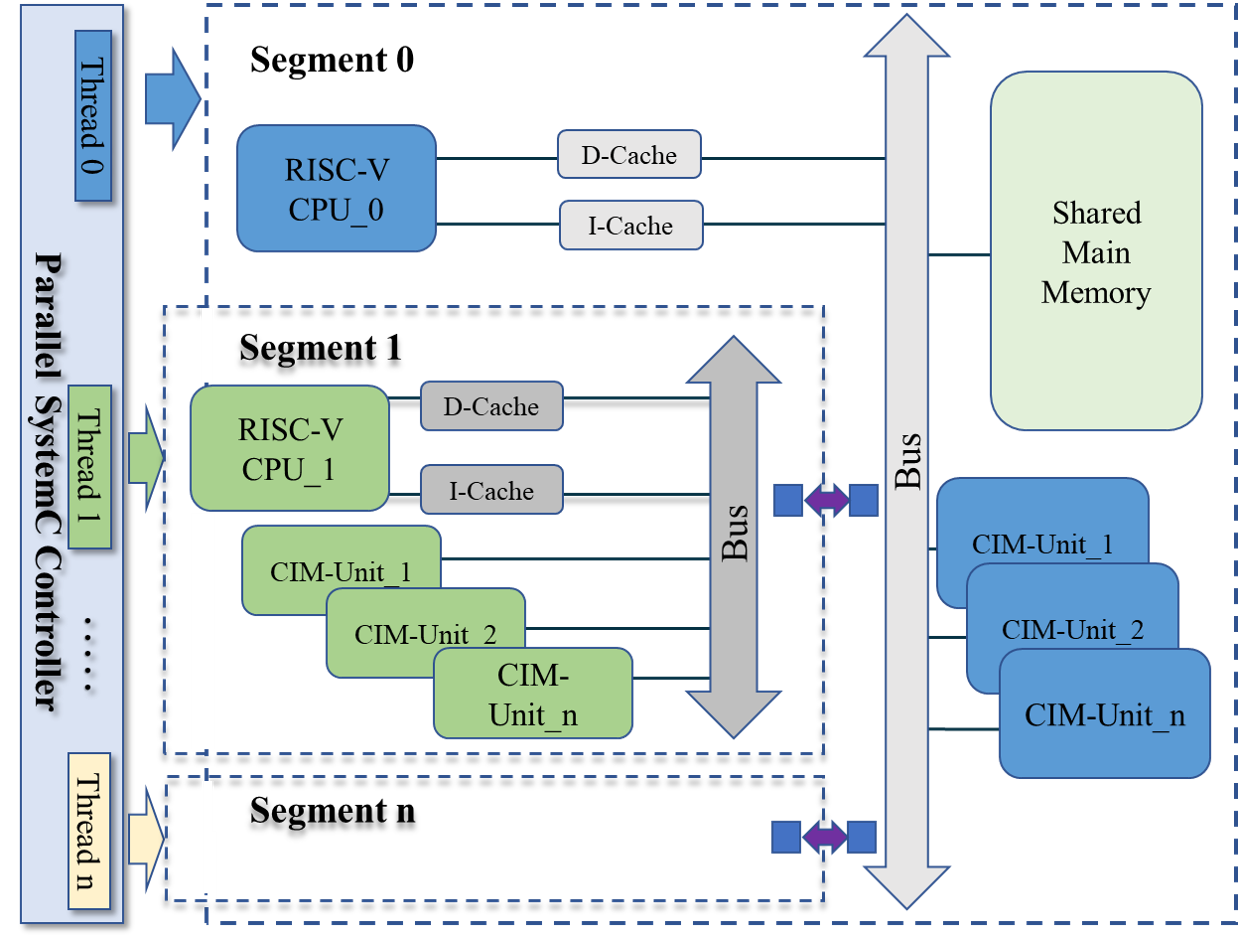}
         \vspace{-5mm}
         \caption{}
         \label{figSg1}
     \end{subfigure}
     \begin{subfigure}[b]{0.45\textwidth}
         \centering
         \includegraphics[width=\textwidth]{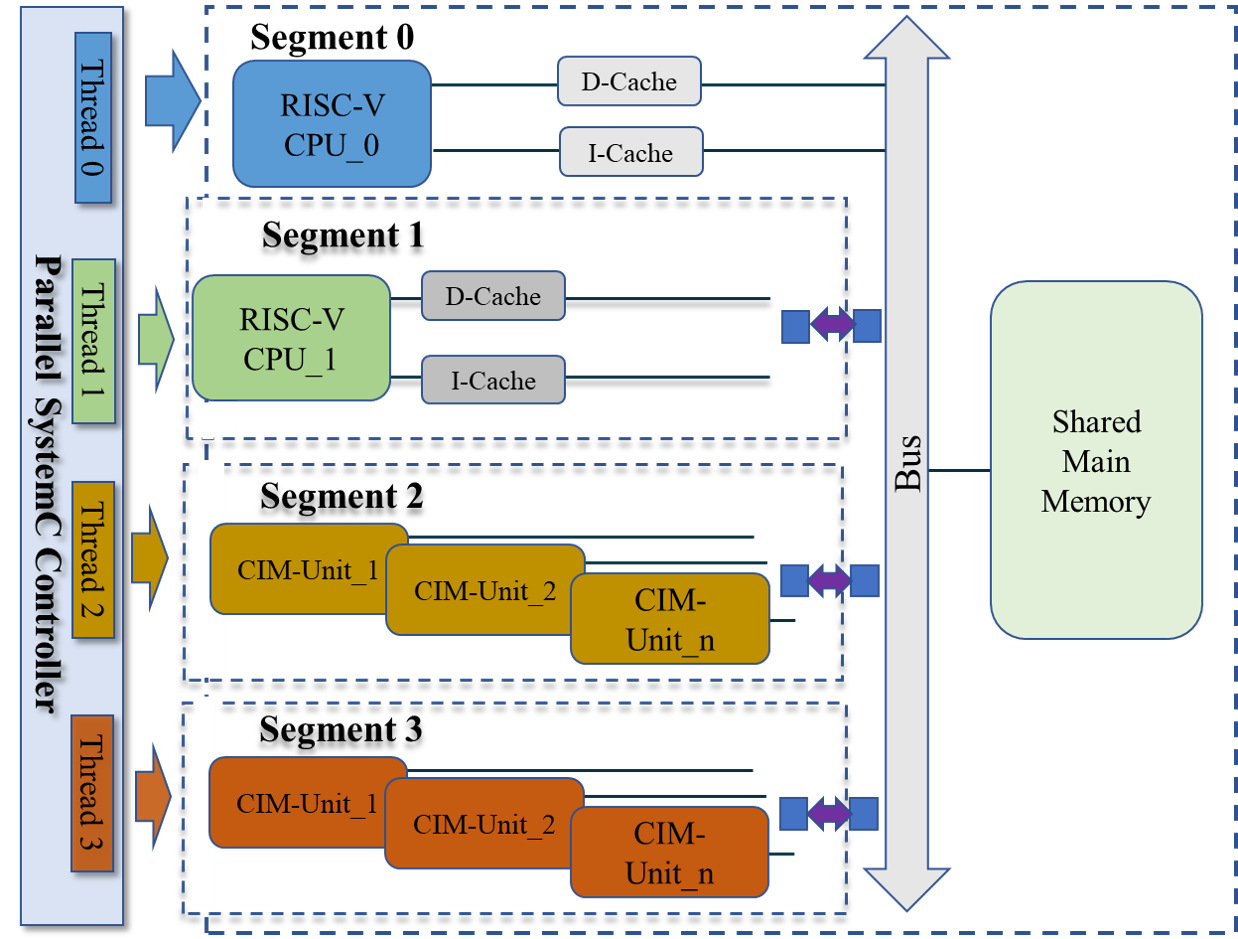}
         \vspace{-5mm}
         \caption{}
         \label{figSg2}
     \end{subfigure}
     \begin{subfigure}[b]{0.45\textwidth}
         \centering
         \includegraphics[width=\textwidth]{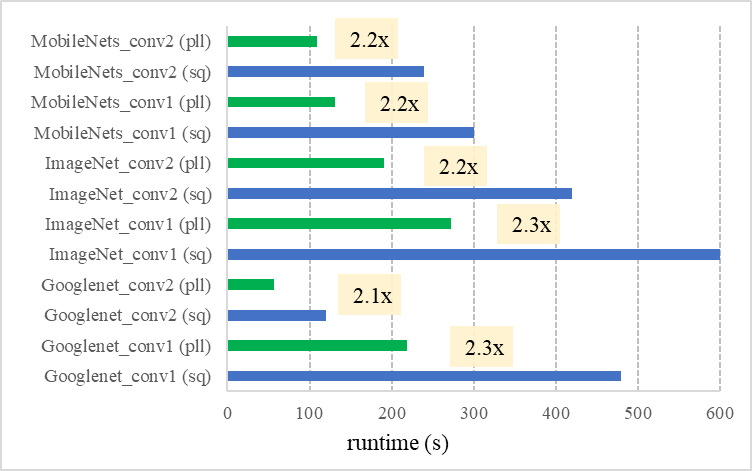}
         \vspace{-5mm}
         \caption{}
         \label{figReSg1}
     \end{subfigure}
     \begin{subfigure}[b]{0.45\textwidth}
         \centering
         \includegraphics[width=\textwidth]{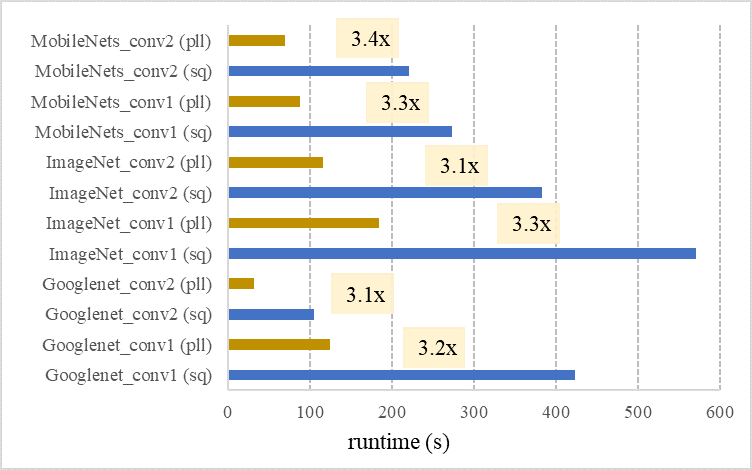}
         \vspace{-5mm}
         \caption{}
         \label{figReSg2}
     \end{subfigure}
    \caption{(a) VP with uniform segmentation, (b) VP with load-oriented segmentation, (c) VMM benchmarks results for VP with uniform segmentation, and (d) VMM benchmarks results for VP with load-oriented segmentation.}
    \vspace{-5mm}
\end{figure*}

\begin{lstlisting}[escapeinside=`']
1 // Simple VMM algorithm: `\begin{math} O = A (h \times w) \cdot B (w \times p) \end{math}'
2  for (  i  = 1  to   `\textbf{h}'   ) {
3     for  (  j  = 1 to `\textbf{p}'  ) {
4       `\begin{math}  O_(i,j)=0 \end{math} '
5         for  ( k  = 1 to  `\textbf{w}'  )  {
6           `\begin{math} O_(i,j)= O_(i,j)+ A_(i,k) \cdot B_(k,j) \end{math} '
7  }  }   }
\end{lstlisting}\label{algo:algo1}



\begin{table}[bp]
\scriptsize
\caption{Multicore VP and Host System Components}
\vspace{-1mm}
\begin{center}
\begin{tabular}{|c|c|}
\hline
\textbf{VP Component} & \textbf{\textit{Parameters}} \\
\hline
RISC-V processor& 2 core 64bits; 1.7GHZ; in order  \\
\hline
L1 I and D caches & SRAM 16KB and 32KB  \\
\hline
Main memory RAM & DRAM 128MB  \\
\hline
CIM-Unit & 2 x Segment ; Crossbar size 256x256  \\
\hline

\textbf{Host Component} & \textbf{\textit{Parameters}} \\
\hline
AMD Ryzen9 processor& 12 core 64bits; 2.2GHZ \\
\hline
L1 I and D caches &  512KB  \\
\hline
Main memory RAM &  61GB  \\
\hline

\end{tabular}
\label{tabcomp}
\end{center}
\end{table}

\subsection{Performance Results}
Results of the simulation runtime of the complete VP systems corresponding to the segmentation illustrated in  Fig.~\ref{figSg1} and  Fig.~\ref{figSg2} are shown in  Fig.~\ref{figReSg1} and in  Fig.~\ref{figReSg2} respectively.  In both cases the VP is executing the benchmarks listed in Table~\ref{tabbench}, once using the conventional sequential (sq) approach and once the parallel simulation (pll) approach. Additionally, for the time-decouple technique the quanta of N instructions was set to 10K after some experimentation. Since increasing the quantum size results in a significant speedup but not uniformly for all benchmarks. When the quantum exceeds the above mentioned value the speed decreases due to the higher number of synchronizations required by the execution of the VMM operations in RISC-V plus main shared memory. When a \textbf{\textit{sync}} is needed all instructions inside the quantum will remain in standby until that \textbf{\textit{sync}} is performed and consequently decreasing the achieved speedup. This effect is not seen when the VMM operations are executed by the CIM-Units due to their extremely low requirement to access new values inside the main memory. Hence, confirming the fact that neuromorphic computing, in this case: computing-in-memory, alleviates the von Neumann bottleneck.

\begin{table}[bp]
\caption{Network Layer Benchmarks}
\vspace{-1mm}
\begin{center}
\scriptsize
\begin{tabular}{|c|c|c|c|c|}
\hline
\textbf{Network Name} & \textbf{\textit{Layer Type id}}& \textbf{\textit{h}}& \textbf{\textit{w}} & \textbf{\textit{p}}\\
\hline
Googlenet& Conv 1 &224&224& 7  \\
\hline
Googlenet& Conv 2 &56&56& 3  \\
\hline
ImageNet& Conv 1 &224&224& 11  \\
\hline
ImageNet& Conv 2 &207&207& 5  \\
\hline
MobileNets& Conv 1 &224&224& 3  \\
\hline
MobileNets& Conv 2 &112&112& 3  \\
\hline

\end{tabular}
\label{tabbench}
\end{center}
\end{table}

From all cases analyzed for the uniform segmentation an up to 2.3× speedup efficiency was registered for ImageNet-conv1 and Googlenet-conv1 network layers. A fast analysis shows that, the total sequential runtime is the sum of segment 0 and 1 individual runtimes as ilustrated in Fig.~\ref{figquantum}. Therefore in this case, it is visible that the obtained speedup is directly related to the number of system host threads performing the simulation, i.e. two parallel segments yield at least 2×  speedup, when compared to sequential, and the rest speedup is attributed to the utilization of the time-decoupled technique. Additionally, from all cases for the load-oriented segmentation an up to 3.3× speedup  efficiency was registered for ImageNet-conv1 and Mobilenets-conv1 network layers.  However in this case, the total speedup relation is not so evident. Again, the total sequential runtime will be the sum of segment 0 to 3 runtimes. Here, from a close examination at the performed load-oriented segmentation, it is posssible to see that segment 1 does not have a heavy load and therefore its contribution to the total speedup is minimal. Nonetheless, the remaining three segments redistribute the total load and yield more than 3× speedup in all benchmarks.

\section{Conclusion}
This work presented a parallel SystemC virtual platform that enables fast simulations for the integration of neuromorphic accelerators within a multicore system. Results show that the utilized simulation approach improves simulation speed by up to 2.3× when a uniform segmentation is performed and up to 3.3× when it is load-oriented for analyzed workloads in both cases respectively. Since current host computers offer multicore computational power to drive several number of threads, exploiting parallel SystemC simulation as a mean to carry out investigation of VPs for neuromorphic systems is shown to be an effective methodology. 

Future work includes the development of automatic segmentation mechanism when the load specifications for the neuromorphic accelerators are known beforehand. Hence, it will allow for a quick starting point to the designer.

\section*{Acknowledgement}
This work was supported by BMBF in Germany within NEUROTEC, project numbers 16ES1134 and 16ES1133K.

\bibliographystyle{ieeetr}
\bibliography{references}

\begin{thebibliography}{10}

\bibitem{neurosurvey2}
F.~Staudigl, F.~Merchant, and R.~Leupers, ``{A Survey of Neuromorphic
  Computing-in-Memory: Architectures, Simulators and Security},'' {\em IEEE
  Design \& Test}, 2021.

\bibitem{Lu1}
A.~Lu, X.~Peng, and S.~Yu, ``{Compute-in-RRAM with Limited On-chip
  Resources},'' in {\em 2021 IEEE 3rd AICAS}, pp.~1--4, 2021.

\bibitem{reram1}
H.~H. Li, Y.~Chen, C.~Liu, J.~P. Strachan, and N.~Davila, ``{Looking Ahead for
  Resistive Memory Technology: A broad perspective on ReRAM technology for
  future storage and computing},'' {\em IEEE Consumer Electronics Magazine},
  vol.~6, no.~1, pp.~94--103, 2017.

\bibitem{vp2}
O.~Bringmann, W.~Ecker, A.~Gerstlauer, A.~Goyal, D.~Mueller-Gritschneder,
  P.~Sasidharan, and S.~Singh, ``The next generation of virtual prototyping:
  Ultra-fast yet accurate simulation of hw/sw systems,'' in {\em DATE}, DATE
  '15, (San Jose, CA, USA), p.~1698–1707, EDA Consortium, 2015.

\bibitem{vp1}
T.~Schmidt, K.~Gr\"{u}ttner, R.~D\"{o}mer, and A.~Rettberg, ``A program state
  machine based virtual processing model in systemc,'' {\em SIGBED Rev.},
  vol.~11, p.~7–12, Jan. 2015.

\bibitem{nvsim}
X.~Dong, C.~Xu, Y.~Xie, and N.~P. Jouppi, ``Nvsim: A circuit-level performance,
  energy, and area model for emerging nonvolatile memory,'' {\em IEEE TCAD},
  vol.~31, no.~7, pp.~994--1007, 2012.

\bibitem{nvmain1}
M.~Poremba and Y.~Xie, ``Nvmain: An architectural-level main memory simulator
  for emerging non-volatile memories,'' in {\em IEEE ISVLSI 2012},
  pp.~392--397, 2012.

\bibitem{mnsim1}
L.~Xia, B.~Li, T.~Tang, P.~Gu, X.~Yin, W.~Huangfu, P.-Y. Chen, S.~Yu, Y.~Cao,
  Y.~Wang, Y.~Xie, and H.~Yang, ``Mnsim: Simulation platform for
  memristor-based neuromorphic computing system,'' in {\em DATE 2016},
  pp.~469--474, 2016.

\bibitem{crosssim}
S.~Agarwal, R.~B. Jacobs~Gedrim, A.~H. Hsia, D.~R. Hughart, E.~J. Fuller, A.~A.
  Talin, C.~D. James, S.~J. Plimpton, and M.~J. Marinella, ``Achieving ideal
  accuracies in analog neuromorphic computing using periodic carry,'' in {\em
  2017 Symposium on VLSI Technology}, pp.~T174--T175, 2017.

\bibitem{neurosim}
P.-Y. Chen, X.~Peng, and S.~Yu, ``Neurosim: A circuit-level macro model for
  benchmarking neuro-inspired architectures in online learning,'' {\em IEEE
  TCAD}, vol.~37, no.~12, pp.~3067--3080, 2018.

\bibitem{nvmspice}
H.~Yu and Y.~Wang, ``Nonvolatile {State} {Identification} and {NVM} {SPICE},''
  in {\em Design {Exploration} of {Emerging} {Nano}-scale {Non}-volatile
  {Memory}} (H.~Yu and Y.~Wang, eds.), pp.~45--83, New York, NY: Springer,
  2014.

\bibitem{aihwkit}
M.~J. Rasch, D.~Moreda, T.~Gokmen, M.~L. Gallo, F.~Carta, C.~Goldberg, K.~E.
  Maghraoui, A.~Sebastian, and V.~Narayanan, ``A flexible and fast pytorch
  toolkit for simulating training and inference on analog crossbar arrays,''
  Apr. 2021.

\bibitem{CIM-SIM}
A.~BanaGozar, K.~Vadivel, J.~Multanen, P.~J{\"a}{\"a}skel{\"a}inen, S.~Stuijk,
  and H.~Corporaal, ``{System Simulation of Memristor Based Computation in
  Memory Platforms},'' in {\em Embedded Computer Systems: Architectures,
  Modeling, and Simulation}, pp.~152--168, Springer International Publishing,
  2020.

\bibitem{Lee2019}
M.~K.~F. Lee, Y.~Cui, T.~Somu, T.~Luo, J.~Zhou, W.~T. Tang, W.-F. Wong, and
  R.~S.~M. Goh, ``A system-level simulator for rram-based neuromorphic
  computing chips,'' {\em ACM Trans. Archit. Code Optim.}, vol.~15, Jan. 2019.

\bibitem{TechnionSystemL}
A.~Eliahu, R.~Ronen, P.-E. Gaillardon, and S.~Kvatinsky, ``{MultiPULPly: A
  Multiplication Engine for Accelerating Neural Networks on Ultra-Low-Power
  Architectures},'' {\em J. Emerg. Technol. Comput. Syst.}, vol.~17, Apr. 2021.

\bibitem{nvmain2}
M.~Poremba, T.~Zhang, and Y.~Xie, ``Nvmain 2.0: A user-friendly memory
  simulator to model (non-)volatile memory systems,'' {\em IEEE Computer
  Architecture Letters}, vol.~14, no.~2, pp.~140--143, 2015.

\bibitem{mnsim2}
Z.~Zhu, H.~Sun, K.~Qiu, L.~Xia, G.~Krishnan, G.~Dai, D.~Niu, X.~Chen, X.~S. Hu,
  Y.~Cao, Y.~Xie, Y.~Wang, and H.~Yang, {\em MNSIM 2.0: A Behavior-Level
  Modeling Tool for Memristor-Based Neuromorphic Computing Systems},
  p.~83–88.
\newblock New York, NY, USA: Association for Computing Machinery, 2020.

\bibitem{ParalleSysC1}
N.~Ventroux, J.~Peeters, T.~Sassolas, and J.~C. Hoe, ``{Highly-parallel
  special-purpose multicore architecture for SystemC/TLM simulations},'' in
  {\em 2014 International Conference on Embedded Computer Systems:
  Architectures, Modeling, and Simulation (SAMOS XIV)}, pp.~250--257, 2014.

\bibitem{OoOparallelsym}
W.~Chen, X.~Han, and R.~Dömer, ``{Out-of-order parallel simulation for ESL
  design},'' in {\em DATE 2012}, pp.~141--146, 2012.

\bibitem{LegaSCI}
C.~Schumacher, J.~H. Weinstock, R.~Leupers, G.~Ascheid, L.~Tosoratto,
  A.~Lonardo, D.~Petras, and T.~Grötker, ``{legaSCi: Legacy SystemC Model
  Integration into Parallel Systemc Simulators},'' in {\em 2013 IEEE
  International Symposium on Parallel Distributed Processing, Workshops and Phd
  Forum}, pp.~2188--2193, 2013.

\bibitem{timedecoup}
J.~H. Weinstock, C.~Schumacher, R.~Leupers, G.~Ascheid, and L.~Tosoratto,
  ``{Time-decoupled parallel SystemC simulation},'' in {\em DATE 2014},
  pp.~1--4, 2014.

\bibitem{SysCArm}
J.~H. Weinstock, R.~L. Bücs, F.~Walbroel, R.~Leupers, and G.~Ascheid,
  ``{Work-in-Progress: AMVP - A High Performance Virtual Platform using
  Parallel SystemC for Multicore ARM Architectures},'' in {\em 2018
  International Conference on Hardware/Software Codesign and System Synthesis
  (CODES+ISSS)}, pp.~1--2, 2018.

\bibitem{ICESOCC}
M.~Galicia, A.~BanaGozar, K.~Sturm, F.~Staudigl, S.~Stuijk, H.~Corporaal, and
  R.~Leupers, ``{NeuroVP: A System-Level Virtual Platform for Integration of
  Neuromorphic Accelerators},'' in {\em IEEE International System-on-Chip
  Conference,to be published}, IEEE SOCC Conference Proceedings, 2021.

\bibitem{Waterman:EECS-2014-54}
A.~Waterman, Y.~Lee, D.~A. Patterson, and K.~Asanović, ``{The RISC-V
  Instruction Set Manual, Volume I: User-Level ISA, Version 2.0},'' Tech. Rep.
  UCB/EECS-2014-54, EECS Department, University of California, Berkeley, May
  2014.

\bibitem{HERDT2020101756}
V.~Herdt, D.~Große, P.~Pieper, and R.~Drechsler, ``{RISC-V based virtual
  prototype: An extensible and configurable platform for the system-level},''
  {\em Journal of Systems Architecture}, vol.~109, p.~101756, 2020.

\bibitem{memCache}
G.~Onnebrink, R.~Leupers, and G.~Ascheid, ``{ESL Black Box Power Estimation:
  Automatic Calibration for IEEE UPF 3.0 Power Models},'' in {\em RAPIDO 2018},
  RAPIDO '18, (New York, NY, USA), Association for Computing Machinery, 2018.

\bibitem{sysClink}
J.~H. Weinstock, R.~Leupers, G.~Ascheid, D.~Petras, and A.~Hoffmann,
  ``{SystemC-link: Parallel SystemC simulation using time-decoupled
  segments},'' in {\em DATE 2016}, pp.~493--498, 2016.

\bibitem{googlenet}
C.~Szegedy, W.~Liu, Y.~Jia, P.~Sermanet, S.~Reed, D.~Anguelov, D.~Erhan,
  V.~Vanhoucke, and A.~Rabinovich, ``Going deeper with convolutions,'' in {\em
  2015 IEEE Conference on Computer Vision and Pattern Recognition (CVPR)},
  pp.~1--9, 2015.

\bibitem{Imagenet}
A.~Krizhevsky, I.~Sutskever, and G.~E. Hinton, ``{ImageNet Classification with
  Deep Convolutional Neural Networks},'' {\em Commun. ACM}, vol.~60,
  p.~84–90, May 2017.

\bibitem{mobilenets}
A.~G. Howard, M.~Zhu, B.~Chen, D.~Kalenichenko, W.~Wang, T.~Weyand,
  M.~Andreetto, and H.~Adam, ``{MobileNets: Efficient Convolutional Neural
  Networks for Mobile Vision Applications},'' {\em CoRR}, vol.~abs/1704.04861,
  2017.

\end{thebibliography}

\end{document}